\documentclass[9pt]{article}
\usepackage{euscript,amssymb}

\catcode`\@=11
\def\lesssim{\mathrel{\mathpalette\vereq<}}
\def\vereq#1#2{\lower3pt\vbox{\baselineskip1.5pt \lineskip1.5pt
\ialign{$\m@th#1\hfill##\hfil$\crcr#2\crcr\sim\crcr}}}

\def\Let@{\relax\iffalse{\fi\let\\=\cr\iffalse}\fi}
\def\vspace@{\def\vspace##1{\crcr\noalign{\vskip##1\relax}}}
\def\multilimits@{\bgroup\vspace@\Let@
 \baselineskip\fontdimen10 \scriptfont\tw@
 \advance\baselineskip\fontdimen12 \scriptfont\tw@
 \lineskip\thr@@\fontdimen8 \scriptfont\thr@@
 \lineskiplimit\lineskip
 \vbox\bgroup\ialign\bgroup\hfil$\m@th\scriptstyle{##}$\hfil\crcr}
\def\Sb{_\multilimits@}
\def\endSb{\crcr\egroup\egroup\egroup}
\def\Sp{^\multilimits@}

\newcommand{\dn}{\mbox{\rm dn}}

\newcommand{\cn}{\mbox{\rm cn}}
\newcommand{\SC}{\mbox{\rm sc}}
\newcommand{\sd}{\mbox{\rm sd}}
\newcommand{\arcsinh}{\mbox{\rm arcsinh}}
\newcommand{\arccosh}{\mbox{\rm arccosh}}
\newcommand{\sign}{\mbox{\rm sign}}
\newcommand{\const}{\mbox{\rm const}}

\begin{document}

\author{\ Mariam Bouhmadi--L\'
opez\footnote{e-mail: mbouhmadi@imaff.cfmac.csic.es}, Pedro F.
Gonz\'alez--D{\'\i}az\footnote{e-mail:
p.gonzalezdiaz@imaff.cfmac.csic.es}, \vspace{+0.5cm}\\ Instituto
de Matem\'aticas y F\'{\i}sica Fundamental, CSIC, \\ C/ Serrano 121,
28006 Madrid, Spain \vspace{+0.5cm}\\ and Alexander
Zhuk\footnote{e-mail: zhuk@paco.net} \vspace{+0.5cm} \\ Department
of Physics, University of Odessa,
\\2 Dvoryanskaya St., Odessa 65100, Ukraine}

\title{Topological defect brane-world models}

\date{ 4 November 2002}

\maketitle

\abstract{ 5-dimensional homogeneous and isotropic models with a
bulk cosmological constant and a minimally coupled scalar field
are considered. We have found that in special cases the scalar
field can mimic a frustrated (i.e. disordered) networks of
topological defects: cosmic strings, domain walls and hyperdomain
walls. This equivalence enabled us to obtain 5-dimensional
instantonic solutions which can be used to construct brane-world
models. In some cases, their analytic continuation to a Lorentzian
metric signature give rise to either 4-dimensional flat or
inflating branes. Models with arbitrary dimensions (D $>$ 5) are
also briefly discussed.}



\section{Introduction}
\setcounter{equation}{0}

For nearly a century, starting with Kaluza--Klein (KK) theories
\cite{Kaluza,Klein} which were seeking for a unification between
gravity and electromagnetism, higher dimensional cosmological
models have been quite often present in scientific research. These
theories has received a great attention during the last years due
to some seminal papers published on brane-worlds, and extra
dimensions models, which are able to provide an explanation for
fundamental physical troubles such as the hierarchy problem. The
geometry as well as the topology of the extra dimensions has to be
compatible with the effective 4--dimensional (4--D) universe where
we live in, i.e. the extra dimension should be hidden: they can be
compact as in the old KK theories or large and infinite as was
firstly pointed out by Rubakov and Shaposhnikov \cite{Ru-Sh},
Akama \cite{Akama} and others \cite{Visser}.

The allowance of more than four dimensions, $4+n$ with $n>0$, for
the space-time introduces a new Planck scale, $M_{PL_{4+n}}$,
which can be considered as a fundamental one and is related to the
usual 4--dimension Planck mass $M_{PL}$. This relation depends on
the topology as well as on the geometry of the extra dimensions
and allows a solution of the hierarchy problem by considering that
$M_{PL_{4+n}}$ is of the order of the weak scale $M_{EW}$
\cite{ADD,RS1}. In the Arkani-Dimopoulus-Dvali (ADD) model
\cite{ADD}, the extra dimensions are compact and
$M_{PL}=M_{PL_{4+n}}^{2+n} R^{n}$, being $R$ the size of the extra
dimensions. The hierarchy problem in this case can be solved
whenever $n\geq 3$ (see e.g. \cite{experiment1}). However, more
recently, Rundall and Sundrum proposed a model (RS1) \cite{RS1}
where turned out to be enough introducing a unique compact extra
dimension. One of the main differences between both models lies on
the dependence of the warp factor on the extra dimension. The RS1
model, inspired by string theory \cite{Witten}, places our
universe in a flat 3--brane, i.e. a 4--D flat hypersurface, with
negative tension, embedded in a 5--D Anti de Sitter (AdS)
space-time, while the second brane with positive tension is
hidden. Later on, Randall and Sundrum constructed another model
(RS2) \cite{RS2} where the extra dimension is infinite and the
universe is lying on a flat 4--D brane with positive tension,
being the bulk a 5--D AdS space-time.

The framework of brane-worlds has been used in cosmology to
describe the birth of branes \cite{GS}. In particular, Garriga and
Sasaki constructed an interesting instanton brane-world model able
to describe the birth of an inflating brane in the semiclassical
approximation. The brane is surrounded by a 5--D AdS space.
However, the general 5--D solution used in this model can present
a singularity when the warp factor vanishes.

In the present paper, we have obtained 5--D solutions which can be
used to construct singularity--free brane-world instantons which,
in some cases, can describe inflating 3-branes. Indeed, we have
obtained a brane-world model where the brane is inflating and the
warp factor has the same dynamical behaviour as in the RS1 and RS2
models \cite{RS1,RS2} (see Eqs.~(\ref{4.13}) and ~(\ref{4.14})
below).

In order to get singularity--free 5--D instantons, we have
considered the bulk filled with matter, which we have modeled by a
minimally coupled scalar field whose energy density and pressure
satisfy a perfect fluid state equation. This scalar field can
mimic the behaviour of different networks of frustrated (i.e.
chaotically distributed) topological defects (NFTD) such as cosmic
strings, domain walls and hyperdomain walls
\cite{Vilenkin84}-\cite{FMP}, depending on the specific state
equation we choose.

The NFTD can be formed during phase transitions. They do not
satisfy scaling solutions and the number of topological defects do
not change in the comoving volume. Up to our knowledge, these
defects were first introduced in cosmology by Kibble, and could
have occasionally dominated the energy density of the early
universe \cite{Kibble}, so changing its dynamical behaviour.
Vilenkin proposed \cite{Vilenkin84} a scenario to describe a
network of frustrated cosmic strings where the strings do not
intercommute, assuming the probability for such physical process
to be nearly zero. The equivalence between such NFTD and scalar
fields has enabled us to obtain 5-dimensional instantonic
solutions which can be used to construct brane-world models.

The paper is organized as follows. In the next section, we
describe the model and get a constraint equation that must satisfy
the scale factor of a 5--D Euclidean homogeneous and isotropic
space endowed with a 5--D cosmological constant and filled with a
scalar field whose energy density and pressure satisfy a perfect
fluid state equation $P=(\alpha -1)\rho$. In section 3, we obtain
the instantonic solutions for $\alpha=3/4$, i.e. when the bulk is
filled with a network of frustrated cosmic strings. Many of these
solutions describe asymptotically AdS wormholes. In section 4, we
get the 5--D Euclidean solutions when the state equation for the
scalar field reads $P=-1/2\rho$. This scalar field can describe a
network of frustrated domain walls. As an example, one of the
solutions (Eq.~(\ref{4.13})) is used to construct a brane-world
instanton which can describe the birth of an inflating brane from
nothing in the semiclassical approximation and whose scale factor
coincides with the one of RS1 and RS2 models. In section 5, we
analyze the case $\alpha=1/4$, which may correspond to a bulk with
a network of frustrated hyperdomain walls (3--D objects). Finally,
we summarize and generalize our model for an arbitrary number of
dimensions ($D> 5$) in section 6.

\section{The model \label{model}}

\setcounter{equation}{0}

As it was stressed before, we are interested in getting 5--D
instanton solutions which can be used to construct brane-world
instantons \cite{GS,BLGDZ} free from any singularity at the origin
of the extra-coordinate. For this purpose, we consider now a 5--D
space filled with a minimally coupled scalar field, $\varphi$, and
a cosmological constant, $\Lambda_5$. The Lorentzian action of
this system has the form
\begin{eqnarray}\label{2.1} S &=& \frac{1}{2\kappa^2_5} \int d^{\,
5}X\sqrt{|g^{(5)}|}\left\{ R[g^{(5)}] - 2 \Lambda_5
\right\}\nonumber \\ &+& \int d^{\, 5}X\sqrt{|g^{(5)}|}\left\{
-\frac12 g^{(5)MN}
\partial_{M} \varphi
\partial_{N} \varphi - V(\varphi ) \right\}
+ S_{YGH} \, ,
\end{eqnarray}
where $\kappa^2_5$ is the 5--D gravitational constant and
$S_{YGH}$ is the boundary term \cite{YGH}. The metric of the
Lorentzian 5--D space $g^{(5)}$ is taken to be homogeneous and
isotropic, so it can be written as
\begin{eqnarray}\label{2.2}
g^{(5)} = g^{(5)}_{MN}dX^M\otimes dX^N = - e^{2\gamma (\tau )}
d\tau \otimes d\tau + e^{2\beta (\tau )} g^{(4)}_{\mu \nu}
dx^{\mu} \otimes dx^{\nu}.
\end{eqnarray}
In this expression, $g^{(4)}$ represents the metric of a 4--D
Einstein space; i.e. $R_{\mu \nu}[g^{(4)}]=\lambda
g_{\mu\nu}^{(4)}$, with constant scalar curvature;
$R[g^{(4)}]=R_4=4\lambda:=12k$ where $k=\pm 1,0$. Unless otherwise
stated, the Einstein spaces are considered as spaces of constant
curvature. However, all the solutions we will obtain are valid for
arbitrary Einstein spaces with constant scalar curvature
$R_4=12k\,\, , k=\pm 1,0$.

Analogously to the case for Eq.~(\ref{2.2}), we consider the
scalar field to be homogeneous. Therefore, its energy density and
pressure are given by:
\begin{eqnarray}\label{2.3} \rho &=& \frac12
e^{-2\gamma} \dot \varphi^2 +V(\varphi )=-T_{0}^{0},\nonumber \\
P&=& \frac{1}{2} e^{-2\gamma} \dot \varphi^2 -V(\varphi
)=T_{\mu}^{\mu}, \quad \mu=1,\dots,4.
\end{eqnarray}
Furthermore, we suppose that the scalar field satisfies the state
equation
\begin{equation}\label{2.4}
P=(\alpha-1)\rho,
\end{equation}
where $\alpha$ is a constant. The conservation equation for the
scalar field energy momentum tensor implies $\rho=A a^{-4\alpha}$,
where $A$ is an arbitrary constant. This results in an inverse
power-law for the scalar field potential $V$ in term of the scale
factor $a= \exp[\beta(\tau)]$; i.e.
$V=(1-\frac{\alpha}{2})A\,a^{-4\alpha}$, whenever $\alpha>0$. This
potential can be rewritten in terms of the scalar field $\varphi$
(see e.g. \cite{Zhuk_QCG}). Therefore, Eq.~(\ref{2.4}) strongly
constrains the form of the potential $V(\varphi)$. Depending on
the values taken on by the parameter $\alpha$, the scalar field
can mimic different kinds of matter as we will discuss shortly.

This model can equivalently be considered as a 5--D space-time
filled with a perfect fluid with action \cite{IM,Zhuk_QCG}
\begin{equation}\label{2.5} S = -\frac{V_4}{\kappa^2_5} \int d \tau
\left[\, 6\, e^{-\gamma + \gamma_0} \dot \beta^2 +e^{\gamma -
\gamma_0} U \right]\, ,
\end{equation}
where $\gamma_0=4\beta$, $V_4 = \int d^4x \sqrt{|g^{(4)}|}$, the
overdot denotes derivative with respect to the Lorentzian time
$\tau$, while the potential $U$ is defined as
\begin{equation}\label{2.6} U = e^{2 \gamma_0} \left( -\frac12
R_4 e^{-2\beta} + \Lambda_5 + \kappa^2_5\, \rho \right).
\end{equation}
{}From Eqs.~(\ref{2.5}), (\ref{2.6}) the scale factor $a$, in the
proper time gauge ($\gamma=0$), must satisfy:
\begin{equation}\label{2.7} \left(\frac{\dot a}{a}\right)^2 +
\frac{k}{a^2} -\frac{\Lambda_5}{6} -\frac{1}{6}\kappa^2_5
A\,a^{-4\alpha} = 0 \, .
\end{equation}

As it has been said before, in this model the matter in the 5--D
space can be equivalently represented by either a scalar field or
a perfect fluid. In both cases, we have to specify the state
equation; i.e. the parameter $\alpha$ given in Eq.~(\ref{2.4}), to
find the dynamical behaviour of the warp factor $a$. There are
many values of $\alpha$ which represent relevant kinds of matter.
For example, if $\alpha=0,1,5/4$, the matter content corresponds
to a 5--D cosmological constant, dust and radiation, respectively.
In this paper, we will consider scalar fields which could mimic a
matter content described by networks of frustrated cosmic strings
(1--D objects) \cite{Vilenkin84}-\cite{BBS}; i.e. $\alpha=3/4$,
domain walls (2--D objects) \cite{BS}-\cite{FMP}; i.e.
$\alpha=1/2$, and also hyperdomain walls (3--D objects); i.e.
$\alpha=1/4$. These topological defects may have been formed
during phase transitions and their evolution afterwards does not
correspond to scaling solutions \cite{VV}. On the other hand, we
shall suppose that the number of topological defects does not
change in the comoving volume. In this paper, we will not study
the underlying field theory that can give rise to such topological
defects.

We analytically continue the constraint equation (\ref{2.7}) into
the Euclidean proper "time" $y$, $\tau\rightarrow-iy$, so that we
will obtain 5--D instantons. In term of $y$ Eq.~(\ref{2.7})
becomes
\begin{equation}\label{2.8}
\left(\frac{da}{dy}\right)^2-k+\frac{\Lambda_5}{6}a^2+
\frac{1}{6}\kappa_5^2 A\,a^{-4\alpha+2}\,=\,0\,.
\end{equation}
{}From the previous equation, it can be seen that for negative
cosmological constant and matter content with $\alpha>0$, the
dominant term for large scale factor ($a\rightarrow+\infty$) is
proportional to $\Lambda_5 a^2$. Therefore, in this limit the
instanton solutions are asymptotically Anti de Sitter (AdS).
Indeed, this behaviour is shown by all instantons for the kind of
matter considered in this work when the warp factor is defined at
large $|y|$ and $\Lambda_5<0$. On the other hand, the instanton
scale factor will obviously depend on the matter content, given by
the cosmological constant and the energy density $\rho$, as well
as the topology of the Euclidean space ($k=\pm 1,0$), as it is
shown by Eq.~(\ref{2.8}). Throughout the paper $r_1$, $a_1$,
$y_1$, $\tau_0$ and $\tau_1$ will mean integration constants.

\section{Cosmic string instantons}

\setcounter{equation}{0}

Networks of frustrated cosmic strings (NFCS) can crop up in a
theory where strings neither intercommute nor pass through each
other \cite{Vilenkin84}. They have been studied in 4--D cosmology
where they could eventually dominate the energy density of the
universe \cite{Kibble}. In this section, we consider that the
energy density $\rho$ in our model behaves like that of a NFCS in
a 5--D space; i.e. $\rho\propto a^{-3}$. The last expression
reminds us the energy density of a 4--D space filled with dust.
The dynamical behaviour of the scale factor is equivalent in both
cases. However, they actually correspond to space-times with
different dimensionality, topology and matter content.

In the time gauge $\gamma=\sqrt{a}\;\; \Rightarrow \;\;
dr:=dy/\sqrt{a}$ and for $\alpha=3/4$, the constraint equation
(\ref{2.8}) becomes:
\begin{equation}\label{3.1}
\left(\frac{da}{dr}\right)^2-ka+\sign(\Lambda)|\Lambda|a^3+\bar{A}^2=0,
\end{equation}
where we have introduced the notation: $\Lambda=\Lambda_5/6$ and
$\bar{A}^2=(1/6)\kappa_5^2 A$. For the particular chosen gauge,
$\gamma=\sqrt{a}$, we can obtain an analytical expression for the
scale factor when the 5--D instanton is sliced into flat,
spherical and hyperbolic sections ($k=\pm 1,0$) and also for
positive, zero and negative cosmological constant.

\subsection{Negative cosmological constant: $\Lambda_5<0$}

The 5--D instantons filled with a scalar field whose energy
density behave like a NFCS, and are 5--D asymptotically AdS
wormholes, when $\Lambda_5$ is negative and the sections
$r=\const$ have constant curvature. The scale factor of these
instantons for $p=\frac{k}{27|\Lambda|^3}+\frac{{\bar
A}^4}{4\Lambda^2}>0$ reads
\begin{eqnarray}\label{3.2}
a(r) = a_1 +
\lambda^2\frac{1-\cn\left(\lambda\sqrt{|\Lambda|}\,(r-r_1)\, |
\,m\right)} {1+\cn\left(\lambda\sqrt{|\Lambda|}\,(r-r_1)\, |
\,m\right)},\quad
0\leq\lambda\sqrt{|\Lambda|}\,|r-r_1|\,\,<\,\,2K(m),
\end{eqnarray}
where
\begin{eqnarray}\label{3.3}
a_1=\left[{\bar A}^2/({2|\Lambda}|)+\sqrt{p}\right]^{1/3}-\;
\left[-{\bar A}^2/({2|\Lambda}|)+\sqrt{p}\right]^{1/3},\quad
\lambda=\left[3a_1^2+\frac{k}{|\Lambda|}\right]^{1/4},\quad
m=\frac{1}{2}-\frac{3a_1}{4\lambda^2}.
\end{eqnarray}
The function $\cn(u | m)$ is a Jacobian Elliptic function while
$K(m)$ is the complete Jacobian Elliptic integral of the first
kind \cite{Abramowitz}. For $r=r_1$, the scale factor is equal to
the minimum radius (throat) of the wormhole, $a_1$. Independently
of the curvature of the 4-D sections ($k=\pm 1,0$), the scale
factor $a$ tends to infinity when
$\lambda\sqrt{|\Lambda|}\,|r-r_1|\rightarrow2K(m)$ (asymptotic AdS
behaviour).

When the parameter $p$ vanishes; i.e. when $\Lambda=-4/(27
\bar{A}^4)$, and the wormhole is sliced into hyperbolic sections,
the solution to equation (\ref{3.1}) becomes
\begin{eqnarray}\label{3.4}
a(r)=a_1\left\{1+\frac{3}{2}\tan^2\left[\sqrt{\frac{3a_1\,
|\Lambda|}{8}}\,(r-r_1)\right]\right\},\quad
0\leq\sqrt{\frac{3a_1|\Lambda|}{8}}\,|r-r_1|\,\,<\,\,\frac{\pi}{2},
\end{eqnarray}
where $a_1$ is defined by Eq.~(\ref{3.3}).

Finally, the scale factor of the instanton for negative values of
$p$ can be written as
\begin{eqnarray}
a(r)=\frac{a_1}{\cn^2\left(\lambda\sqrt{|\Lambda|}\,(r-r_1)\, |
\,m\right)} - a_3\;\SC^2\left(\lambda\sqrt{|\Lambda|}\,(r-r_1)\, |
\,m\right) ,\quad
0\leq\lambda\sqrt{|\Lambda|}\,|r-r_1|\,\,<\,\,K(m),
\label{3.5}\end{eqnarray}
where the function $\SC (u | m)$ again is a Jacobian Elliptic
function and
\begin{eqnarray} a_1&=&2s^{1/3}\cos(\theta/3),
a_2=-s^{1/3}[\cos(\theta/3)+\sqrt{3}\sin(\theta/3)],
a_3=s^{1/3}[-\cos(\theta/3)+\sqrt{3}\sin(\theta/3)],\nonumber\\
s&=&\left(\frac{1}{27|\Lambda|^3}\right)^{1/2},\quad
\theta=\arctan\left[\frac{2|\Lambda|\sqrt{-p}}{\bar{A}^2}\right],\quad
\lambda =\frac{1}{2}(a_1-a_2)^{1/2},\quad
m=\frac{a_3-a_2}{a_1-a_2}. \label{3.6}\end{eqnarray}
The parameter $\theta$ takes values on the interval $(0,\pi/2)$.
This can be deduced by taking into account that
$\cos(\theta)=\bar{A}^2/(2|\Lambda| s)$ and
$\sin(\theta)=\sqrt{-p}/s$. On the one hand, the negativeness of
$p$ in the case under consideration forbids the value $\theta=0$.
On the other hand, the existence of a non zero matter content in
the model ($\bar{A}\neq0$) does not allow the value
$\theta=\pi/2$.

The cases described by Eqs. (\ref{3.4}) and (\ref{3.5}) represent
also Euclidean asymptotically AdS wormholes. In fact, the scale
factor varies from $a_1$ ($a_1=a(r_1)$), corresponding to the
radius of the wormhole throat, up to infinity as
$\sqrt{{3a_1|\Lambda|}/{8}}\,|r-r_1|\rightarrow{\pi}/{2}$ or
$\lambda\sqrt{|\Lambda|}\,|r-r_1|$ approaches $K(m)$ for
(\ref{3.4}) and (\ref{3.5}), respectively.

All the above solutions can be continued back into the Lorentzian
time ($(r-r_1)\rightarrow i(\tau-\tau_1)/\sqrt{a}$). In this case,
it can be seen that the Lorentzian solutions represent 5--D
collapsing FRW universes as the scale factors then vary between 0
and $a_1$ where the explicit expression of $a_1$ depends on the
type of 4--d geometry ($k=\pm 1,0$) and the parameter $p$.

\subsection{Positive cosmological constant: $\Lambda_5>0$}

Unlike for negative cosmological constant, there are not Euclidean
solutions for $r=\const$ with hyperbolic ($k=-1$) or flat ($k=0$)
geometry, when the 5--D space is filled with NFCS and positive
cosmological constant. This is because in these cases the
differential equation (\ref{3.1}) would imply
$(\frac{da}{dr})^2<0$.

It will however be seen that there are instantonic solutions with
spherical $r=\const$ sections. The explicit form of such solutions
(for $k=1$) depends on the parameter $p=-\frac{1}{27
\Lambda^3}+\frac{{\bar A}^4}{4\Lambda^2}$. If $p\geq0$, then there
are not Euclidean solution because Eq.(\ref{3.1}) would then imply
$\left(\frac{da}{dr}\right)^2<0$. The situation is rather
different for $p<0$, as in this case there is an Euclidean 5--D
space whose scale factor $a$ is
\begin{eqnarray}
a(r)= \frac{a_3}{\dn^2\left(\lambda\sqrt{\Lambda}(r-r_1)\, |
\,m\right)} -a_2\;m\;\sd^2\left(\lambda\sqrt{\Lambda}(r-r_1)\, |
\,m\right),\quad 0\leq\lambda\sqrt{\Lambda}|r-r_1|\,\,\leq\,\,
K(m).\label{3.7}
\end{eqnarray}
The functions $\dn (u | m)$ and $\sd (u | m)$ again are Jacobian
Elliptic functions, the parameters $\lambda, a_1$, $a_2$, $a_3$
$s$, $\theta$ are defined in Eq.~(\ref{3.6}) with the obvious
substitution $|\Lambda|$ by $\Lambda$, and
\begin{equation}
m=\frac{a_1-a_3}{a_1-a_2}.\label{3.8}
\end{equation}

The warp factor of this instanton takes on values between $a_3$
for $r=r_1$ and $a=a_1$ for $\lambda\sqrt{\Lambda}|r-r_1|=K(m)$.
The parameter $\theta$ takes on values in the interval
$(\frac{\pi}{2},\pi)$ as $\cos(\theta)=-\bar{A}^2/(2\Lambda s)$
and $\sin(\theta)=\sqrt{-p}/s$. The presence of matter in the
model ($\bar{A}\neq 0$) does not allow the case $\theta=\pi/2$,
while the negativeness of $p$ ($p<0$) forbids the value
$\theta=\pi$. In the semiclassical approximation, this instanton
describes the tunneling between a collapsing 5--D FRW universe
with a maximum radius equal to $a_3$ and an asymptotically de
Sitter (dS) space-time with a minimum scale factor $a=a_1$.

\subsection{Zero cosmological constant: $\Lambda_5=0$}

If $\Lambda_5$ vanishes, there is a unique non-trivial instantonic
solution with the topology $\mathbb{R}\times\mathbf{S}^4$ and the
following scale factor:
\begin{equation}\label{3.9}
a(r)=\bar{A}^2+\frac{1}{4}(r-r_1)^2, \quad r\in\mathbb{R}.
\end{equation}
This instanton is a 5--D asymptotically flat wormhole whose throat
is located at $r=r_1$. It is worth mentioning that a fine tuning
between the matter content and the spatial curvature:
$ka=\bar{A}^2 \,\Longrightarrow\, a=a_1:=\bar{A}^2$, leads to a
"static trivial solution" with constant scale factor, $a_1$, and
topology $\mathbb{R}\times\mathbf{S}^4$. This static instanton is
unstable.

There is a baby universe branching off from the wormhole throat.
Bearing in mind that the energy density of a NFCS in a 5--D space
scales respect to the warp factor $a$ similarly to as dust did in
a 4--D space, the behaviour of this baby universe is similar to
the dynamics of a 4--D closed FRW universe filled with dust.

\section{Domain wall instantons}

\setcounter{equation}{0}

Domain walls may have been formed during phase transition when the
vacuum manifold has disconnected components. Particularly
interesting are the networks of frustrated domain walls (NFDW)
\cite{BS} whose evolution does not approach a scaling solution.
They have been sketched recently in the literature to give account
of the dark matter in the universe \cite{BBS} as well as the dark
energy \cite{FMP}. While in a 4--D universe the energy density of
a NFDW is proportional to $a^{-1}$, in a 5--D space it behaves as
$a^{-2}$. Therefore, the presence of a scalar field in a 5--D
space with an energy density similar to the one of a NFDW implies
the appearance of an effective curvature term. This has
interesting consequences as we will see shortly.

\subsection{Negative cosmological constant: $\Lambda_5<0$}

We are now going to describe 5--D Euclidean spaces filled with a
matter content with state equation $P=-1/2\rho$. The matter
content can either be modeled by a scalar field or a perfect
fluid. Both cases can describe effectively a NFDW. For convenience
we rewrite the constraint equation (\ref{2.8}) (for $\alpha=1/2$)
in the proper Euclidean time gauge as
\begin{equation}\label{4.10}
\left(\frac{da}{dy}\right)^2+\sign(\Lambda)|\Lambda| a^2
-k_{eff}=0.
\end{equation}
In this equation, $k_{eff}=k-\bar{A}^2$ is the effective spatial
curvature parameter. This parameter encodes the geometry of the
$y=\const$ sections by means of $k$ and the matter content through
$\bar{A}^2$. For a negative cosmological constant, the solution to
Eq.~(\ref{4.10}) reads
\begin{equation}\label{4.11}
a(y)=\sqrt{\left|\frac{k_{eff}}{\Lambda}\right|}
\cosh\left[\sqrt{|\Lambda|}(y-y_1)\right], \quad y\in\mathbb{R},
\end{equation}
when $k_{eff}$ is negative ($k_{eff}<0$). This instanton describes
a 5--D asymptotically AdS wormhole whose throat is located at the
hypersurface $y=y_1$. Even if the behaviour of the scale factor of
the wormhole is similar for all negative values of the parameter
$k_{eff}$, its topology can be different as it can be sliced into
flat, hyperbolic or spherical $y=\const$ sections. Indeed, for a
given 5--D cosmological constant, the radius of the wormhole
throat get narrower, when the instanton is sliced into spherical
4--D hypersurfaces, than in hyperbolic 4--D ones.

The behaviour of the warp factor $a$ is different for strictly
positive values of $k_{eff}$. In this case
\begin{equation}\label{4.12}
a(y)=\sqrt{\frac{k_{eff}}{|\Lambda|}}\,\sinh\left[\sqrt{|\Lambda|}\,
|y-y_1|\right]\, ,\quad y\in\mathbb{R},
\end{equation}
while the topology of the instanton is
$\mathbb{R}\times\mathbf{S}^4$. In contrast with the solution
described by Eq.~(\ref{4.11}), the scale factor vanishes at
$y=y_1$ and could induce a singularity at this hypersurface
($y=y_1$). It happens when $k_{eff}<1$, because in this case the
scalar curvature blows up when $y$ approaches $y_1$. However, for
$k_{eff}=1$ (absence of matter), the solution (\ref{4.12})
describes an Euclidean AdS space when the $y=\const$ sections are
spherical and the scalar curvature is well-defined at the origin
of the extra-coordinate $y=y_1$. The situation can be rather
different for instantons whose scale factor equals solution
(\ref{4.12}) with $k_{eff}=k=1$, sliced into more general 4--D
Einstein spaces with $R[g^{(4)}]=12$. In the later case, this is
so because the invariant $C^2$ related to the 5--D Weyl tensor of
the 5--D instanton by $C^2\equiv {C^{\mu}}_{\nu\rho\sigma}
{C_{\mu}}^{\nu\rho\sigma}$ can be divergent at the origin of the
extra-coordinate. For $k_{eff}=1$, the instanton solution
(\ref{4.12}) has been used to construct a brane-world model
\cite{GS}.

Finally, for vanishing $k_{eff}$, i.e. when the 5--D instanton has
spherical sections at $y=\const$ and a matter content with
$\bar{A}^2=1$, the warp factor is:
\begin{equation}\label{4.13}
a(y)=a_1\exp(\pm\sqrt{|\Lambda|}\, y\,),\;\;\; a_1>0, \quad
y\in\mathbb{R}.
\end{equation}
This scale factor coincides with the one of an Euclidean 5--D AdS
space sliced into 4--D flat spatial sections. However, an
instanton, with the scale factor (\ref{4.13}), sliced into 4--D
spherical sections is not an Euclidean 5--D AdS. In fact, its
scalar curvature depends on the Euclidean extra coordinate , $y$,
\begin{eqnarray}
R[\,g^{(5)}\,]= \frac{12}{a_1^2}\exp(\mp\,2\sqrt{|\Lambda|}\,y\,)
-20|\Lambda|, \nonumber
\end{eqnarray}
and get asymptotically AdS for large values of $a(y)$. It is clear
that solution (\ref{4.13}) with the fine tuning $\bar{A}^2=1$ is
unstable: any matter fluctuations $\bar{A}^2 + \delta \bar{A}^2$
will result in transition of (\ref{4.13}) into either (\ref{4.11})
or (\ref{4.12}), depending on the sign of $\delta \bar{A}^2$, with
$k_{eff} = \delta \bar{A}^2$ and the same topology
$\mathbb{R}\times\mathbf{S}^4$.

Instanton (\ref{4.13}) is nonsingular at any finite value of
$|y|$. It can be used to construct a brane-world instanton
\cite{GS,BLGDZ}. For example, this can be done, firstly, by
excising from the instanton the regions: $y>0$ for \mbox{$a=a_1
\exp(\sqrt{|\Lambda|}y)$}, and, $y<0$ for \mbox{$a=a_1
\exp(-\sqrt{|\Lambda|}y)$}, and secondly, by gluing the remaining
instanton pieces at $y=0$. The resulting solution corresponds to a
brane-world instanton with a 4--D spherical brane located at
$y=0$, with tension $T_1\,=\,6/\kappa_5^2\sqrt{|\Lambda|}$.

The above brane-world instanton can describe the birth of
inflating brane. In fact, it can be interpreted as a semiclassical
path for the quantum tunneling from nothing (the Euclidean
region). This can be easily seen by performing an analytical
continuation along the azimuthal coordinate $\chi$ of the 4--D
sphere: \mbox{$\chi\,\rightarrow \,iHt+{\pi}/{2}$}. The resulting
Lorentzian metric reads:
\begin{equation}\label{4.14} ds^2_L = dy^2 +  H^{2}a^2(y)( - dt^2 +
\frac{1}{H^2}\cosh^2 Ht \; d\Omega^2_{(3)})\, ,
\end{equation}
where $a(y)=a_1 \exp(-\sqrt{|\Lambda|}\,|y|)$. Here, the parameter
$H$ is chosen in such a way that $t$ describes the proper time on
the brane; i.e $H=1/a(y)|_{y=0}=1/a_1$. The metric (\ref{4.14})
describes an inflating brane located at $y=0$ with a positive
tension $T_1$. This brane-world has a warp factor similar to those
of the RS1 and RS2 models \cite{RS1,RS2}. However, in our model
the brane is inflating , it is a 4--D de Sitter space-time (dS)
with Hubble constant $H$, instead of being static and flat as in
\cite{RS1,RS2}.

The others 5--D Euclidean solutions described in the present paper
can also be used to construct brane-world instantons by a similar
procedure of cutting and gluing of the original instanton
\cite{GS,BLGDZ}. Some of this gravitational instantons with branes
can describe the creation of brane-worlds.

\subsection{Positive cosmological constant: $\Lambda_5>0$}

For a space filled with a scalar field, with $P=-1/2\rho$, and a
positive cosmological constant, there is an unique Euclidean
solution when the $y=\const$ sections have constant scalar
curvature. The solution to Eq.~(\ref{4.10}) reads
\begin{eqnarray}
a(y)=\sqrt{\frac{k_{eff}}{\Lambda}}\sin\left[\sqrt{\Lambda}\,
|y-y_1|\right],\quad
0\leq\sqrt{\Lambda}|y-y_1|\,\,\leq\,\,\frac{\pi}{2},\label{4.15}
\end{eqnarray}
and the effective spatial curvature parameter must be $0<k_{eff}
\leq 1$; i.e. $k=1$ and $\bar{A}^2<1$. In the limiting case
$k_{eff}=1$ (absence of matter), the instanton describes a 5--D
sphere, that equivalently is an Euclidean dS space.

In the semiclassical approximation, solution (\ref{4.15})
describes a quantum path for the creation of a 5--D asymptotically
\mbox{dS space-time} from nothing. However, similarly to solution
(\ref{4.12}), this instanton is singular at the hypersuface
$y=y_1$ when $k_{eff}<1$, because the scalar curvature reaches
infinite value when $y$ approaches $y_1$.

\subsection{Zero cosmological constant: $\Lambda_5=0$}

If the 5--D cosmological constant vanishes, the scale factor
solution of Eq.~(\ref{4.10}) is
\begin{eqnarray}\label{4.16}
a(y)=a_1\,\pm\,\sqrt{k_{eff}}\;(y-y_1)\;,\quad
-{a_1}\leq\;\pm\;{\sqrt{k_{eff}}}(y-y_1),
\end{eqnarray}
while the instanton topology corresponds to
$\mathbb{R}\times\mathbf{S}^4$ and $0\,\leq\, k_{eff}\,\leq\, 1$.
When the effective curvature parameter equals unity (absence of
matter), the previous solution describes a flat 5--D Euclidean
space sliced into 4--D spherical sections. On the other hand, if
$k_{eff}=0$, the instanton has an arbitrary constant warp factor
which can have a runaway behaviour similar to Einstein universe.
Therefore, in this case, the solution is unstable\footnote{A small
perturbation on the effective spatial curvature parameter leads to
a linear growth of the instanton scale factor.}. For the remaining
cases, $0<k_{eff}<1$, the instanton is asymptotically flat and has
a singularity at the origin of the extra dimension $y=y_1\,\mp
a_1/\sqrt{k_{eff}}$.

\section{Hyperdomain wall instantons}

\setcounter{equation}{0}

Since we are working in 5--D spaces, there can still be another
kind of topological defects which might have been formed during
early phase transition. We have called them hyperdomain walls and
they are 3--D objects. Analogously, to strings and domains walls,
these topological defects could give rise to networks of
frustrated hyperdomain walls (NFHW) depending on the underlying
broken symmetry. The energy density of a NFHW scales as $a^{-1}$
and can also be effectively modeled either by a scalar field or by
a perfect fluid, with the state equation $P=-3/4\,\rho$ (this
corresponds to $\alpha=1/4$ in Eq.~(\ref{2.4})). A 5--D universe
filled with this matter content has a dynamical behaviour similar
to a 4--D universe filled with NFDW, as the energy density in both
cases has the same behaviour respect to the scale factor.

\subsection{Negative cosmological constant: $\Lambda_5<0$}

We can rewrite the constraint equation (\ref{2.8}) in the proper
Euclidean time gauge for $\alpha=1/4$ as follows:
\begin{equation}\label{5.1}
\left(\frac{da}{dy}\right)^2+\sign(\Lambda)|\Lambda| a^2
+\bar{A}^2 a -k=0.
\end{equation}
If $\Lambda_5<0$ and the 5--D space is filled with a scalar field
with $P=-3/4\,\rho$, there is a solution which describes Euclidean
asymptotically AdS wormholes whose scale factor reads
\begin{eqnarray}\label{5.2}
a(y)= \frac{1}{2\,|\Lambda|} \left\{\bar{A}^2 +
\sqrt{q}\cosh[\sqrt{|\Lambda|}\,(y-y_1)] \right\}, \quad
y\in\mathbb{R}.
\end{eqnarray}
Here, the parameter $q$, defined as $q:=\bar{A}^4-4\,k\,|\Lambda|$
and $k=\pm 1, 0\,$; is strictly positive. This wormhole can be
sliced into flat, hyperbolic or spherical 4--D sections and its
throat radius, $a_1=a(y_1)$, located at $y=y_1$ is smaller when
$k=1$. There is a collapsing baby universe branching off from the
throat of the wormhole. When the 4--D geometry is flat or
hyperbolic, the scale factor of the baby universe takes values on
$[0,a_1]$, while for $k=1$, the baby universe has a minimum radius
different from zero $a_0=(\bar{A}^2-\sqrt{q})/({2|\Lambda|})$. In
this case, $q>0$ and $k=1$, the Lorentzian metric of the universe
interpolates between the wormhole solution (\ref{5.2}), with
$k=1$, and the instanton\footnote{This Lorentzian metric has the
form of equation (\ref{2.2}), with $\gamma=0$, where the scale
factor $a$ is obtained either from (\ref{5.2}) or from (\ref{5.3})
with the help of the Wick rotation: $y-y_1\to i(\tau-\tau_1)$ or
$y-y_0\to i(\tau-\tau_0)$, i.e.,
$a(\tau)={1}/({2|\Lambda|})\left\{\bar{A}^2-\sqrt{q}
\cos\left[\sqrt{|\Lambda|}|\tau-\tau_0|\right]\right\}$,
$0\leq|\tau-\tau_0|\leq\pi/\sqrt{|\Lambda|}$.}
\begin{eqnarray}\label{5.3}
a(y)=\frac{1}{2|\Lambda|}\left\{\bar{A}^2-\sqrt{q}
\cosh\left[\sqrt{|\Lambda|}\,|y-y_0|\right]\right\},\quad
0\leq\sqrt{|\Lambda|}\,\,|y-y_0|\leq
\arccosh\left[\frac{\bar{A}^2}{\sqrt{q}}\right].
\end{eqnarray}
The scale factor (\ref{5.3}) vanishes at
$y=\arccosh(\bar{A}^2/\sqrt{q})$ and reaches its maximum value,
$a_0$, at $y=y_0$, where $y_0$ is an integration constant.

When the parameter $q=0$, i.e. $\bar{A}^4=4k|\Lambda|$, there are
5--D instanton solutions only for $k=1$, and the scale factor
reads
\begin{equation}\label{5.4}
a(y)=\left(a_1-\frac{1}{\sqrt{|\Lambda|}}\right)\exp\left[
\sqrt{|\Lambda|}\,(y-y_1)\right]+ \frac{1}{\sqrt{|\Lambda|}}\,.
\end{equation}
On the one hand, if $a_1>1/\sqrt{|\Lambda|}$,
$\sqrt{|\Lambda|}\,(y-y_0)\,\in\,\mathbb{R}$, the scale factor is
larger than $1/\sqrt{|\Lambda|}$ and approaches
$1/\sqrt{|\Lambda|}$, when
\mbox{$\sqrt{|\Lambda|}(y-y_1)\rightarrow -\infty$}. This solution
is asymptotically AdS (it can be easily checked, that scalar
curvature $R \to -20|\Lambda|$ as $a(y) \to +\infty$). On the
other hand, if \mbox{$a_1<1/\sqrt{|\Lambda|}\,$},
$\sqrt{|\Lambda|}\,(y-y_1)\,\in\,(-\infty,
-\ln\{1-a_1\sqrt{|\Lambda|}\,\}\,]$ and the scale factor can take
on values in the interval $[\,0,{1}/{\sqrt{|\Lambda|}}\,)$. When
\mbox{$\sqrt{|\Lambda|}(y-y_1)=-\ln(1-a_1\sqrt{|\Lambda|})$}, $a$
vanishes and the scalar curvature diverges, while for
\mbox{$\sqrt{|\Lambda|}(y-y_1)\rightarrow -\infty$}, $a$
approaches $1/{\sqrt{|\Lambda|}}$. In both cases, it is impossible
to perform an analytical continuation to a Lorentzian metric whose
proper time would be $\tau$. This can be easily seen by checking
that the extrinsic 5--D curvature does not vanish for any finite
value of the Euclidean proper time $y$. It must be noted that
there is a "static solution", where the scale factor takes a very
particular value $a=1/\sqrt{|\Lambda|}$. Similar to the 4-D static
Einstein universe, this instanton is unstable: a small variation
of the cosmological constant or the matter content (through
$\bar{A}$), leads to an exponential increasing or decreasing of
the scale factor (with $y$).

Finally, when the parameter $q$ is strictly negative, there is a
5--D Euclidean solution whose scale factor is
\begin{eqnarray}
a(y)=\frac{1}{2|\Lambda|}\left\{\bar{A}^2 + \sqrt{-q}\sinh\left[
\sqrt{|\Lambda|}(y-y_1)\right] \right\},\quad
\sqrt{|\Lambda|}(y-y_1) \ge
-\arcsinh\left[\frac{\bar{A}^2}{\sqrt{-q}}\right]. \label{5.5}
\end{eqnarray}
The topology of the previous instanton is
\mbox{$\mathbb{R}\times\mathbf{S}^4$} and it is asymptotically
AdS. The scale factor can take any positive value. Moreover, it
vanishes when
$\sqrt{|\Lambda|}(y-y_1)=-\arcsinh\left[{\bar{A}^2}/{\sqrt{-q}}
\right]$, where the scalar curvature diverges and consequently the
instanton is singular at that 4--D hypersurface.

\subsection{Positive cosmological constant: $\Lambda_5>0$}
It can be easily seen that for positive bulk cosmological
constant, the Euclidean equation (\ref{5.1}) has a unique
instanton solution which is sliced into 4--D spherical $y=\const$
sections and whose scale factor reads
\begin{eqnarray}\label{5.6}
a(y)=\frac{1}{2\Lambda}\left\{-\bar{A}^2+\sqrt{\bar{A}^4+4\Lambda}
\cos\left[\sqrt{\Lambda}(y-y_1)\right]\right\}, \quad 0\,\leq\,
\sqrt{\Lambda}\,|y-y_1|\,\leq\,\arccos\left(\frac{\bar{A}^2}
{\sqrt{\bar{A}^4+4\Lambda}}\right).
\end{eqnarray}
The scale factor of this instanton solution vanishes when
$\sqrt{\Lambda}|y-y_1|=\arccos({\bar{A}^2}/{\sqrt{\bar{A}^4+4\Lambda}})$.
Moreover, the vanishing of the scale factor induces a singularity
as the scalar curvature diverges in this hypersurface. It must be
also noted that there is a 5--D Lorentzian universe branching off
from this instantonic solution at $y=y_1$. In fact, if we perform
the analytical continuation $\sqrt{\Lambda}(y-y_1)\rightarrow
i\sqrt{\Lambda}(\tau-\tau_1)$, we obtain a 5--D asymptotically dS
universe.

\subsection{Zero cosmological constant: $\Lambda_5=0$}

Analogously to the positive cosmological constant case, the
constraint equation (\ref{5.1}) has a unique non-trivial solution
when $\Lambda_5$ is zero\footnote{A fine tuning between the matter
content and the spatial curvature: $k=\bar{A}^2 a$, leads to a
"static trivial solution" with constant scale factor, $a=\bar
A^{-2}$, and 4--D spherical $y=\const$ sections. This solution is
unstable. For example, small fluctuations around its Lorentzian
counterpart will grow with time quadratically.}. The scale factor
of the instanton is
\begin{eqnarray}\label{5.7}
a(y)=\frac{1}{\bar{A}^2}\left[1-\frac{\bar{A}^4}{4}(y-y_1)^2\right],
\quad 0\,\leq\,|y-y_1|\leq\,\frac{2}{\bar{A}^2},
\end{eqnarray}
and has the topology $\mathbb{R}\times\mathbf{S}^4$. This
Euclidean solution has a singularity at $|y-y_1|=2/{\bar{A}^2}$,
where the scale factor vanishes.There is a 5--D universe branching
off from the hypersurface corresponding to the maximum radius of
the scale factor, i.e. $y=y_1$.

\section{Conclusions}

\setcounter{equation}{0}

In the present paper we have described 5--D gravitational
instantons whose geometry is homogeneous and isotropic in a space
which is filled with a cosmological constant, $\Lambda_5$, and a
minimally coupled scalar field, $\varphi$, that satisfies a
perfect fluid state equation. In particular, we have chosen the
5--D space to be sliced into 4--D constant curvature spaces.
However, all our solutions are valid when the $y=\const$ sections
are 4--D Einstein spaces with constant scalar curvature. The
scalar field potential, $V(\varphi)$, is strongly constrained by
the perfect fluid state equation $P=(\alpha-1)\rho$. Indeed, we
have seen that in terms of the scale factor $a$, $V(\varphi)$
satisfies an inverse power law, whenever $\alpha>0$.

The matter content in the model can equivalently be represented
either by a minimally coupled scalar field or by a perfect fluid
with a given state equation. In both cases, it can be noticed that
in the Euclidean sector, the energy density $\rho$ is frozen. On
the other hand, we have seen that $\varphi$ can mimic different
types of networks of frustrated topological defects, including
cosmic strings ($\alpha=3/4$), domain walls ($\alpha=1/2$) and
hyperdomain walls ($\alpha=1/4$). For all of these equations of
state and for the different possibilities of the cosmological
constant (positive, negative and zero) we have got the behaviour
of the scale factor $a$. These instantons are asymptotically AdS
when $\Lambda_5$ is negative and the scale factor takes on large
values ($a\rightarrow +\infty$). Moreover, we have checked that
some of the Euclidean solutions are wormholes.

The obtained instantons are useful to construct nonsingular
brane-world models, such as we have explicitly done in section 4
for solution (\ref{4.13}). There, we have seen that an Euclidean
space with a negative cosmological constant and a NFDW matter
content when $k_{eff}=0$, i.e. $\bar{A}=1$ and $k=1$, has a scale
factor similar to those of RS1 and RS2 models \cite{RS1,RS2}. This
space can be used to build a brane-world instanton by a cutting
and gluing procedure. Some of the advantages of this model are
that it is not singular and can describe the birth of an inflating
brane in the semiclassical approximation. Here, inflation is due
only to the geometrical construction. Since the sections of
constant Euclidean "time" ($y=\const$) are spherical, the
analytical continuation along the azimuthal coordinate, $\chi$, of
the 4--D sphere, $\chi\rightarrow iHt+\pi/2$, results in a brane
whose geometry is a 4--D dS space-time. Here, the bulk energy
density $\rho$ remains frozen after the analytical continuation.

The cutting and gluing procedure can also be used to construct
models with an arbitrary number of parallel branes \cite{BLGDZ}.
These branes can either be inflating or static, after their birth,
whenever the topology of the $y=\const$ are $\textbf{S}^{4}$ or
$\mathbb{R}^{4}$, respectively. It must be noted that we can also
use instantons with a singularity at some hypersurface $y=\const$,
by introducing an additional brane and removing from the
brane-world instanton the singular region. This could lead to the
appearance of branes with negative tension.

It is worth remarking that model (\ref{2.2}) can be generalized
for an arbitrary number of dimensions: $g^{(4)} \to g^{(d)}$ and $
\mbox{$\mathbb{R}\times\mathbf{M}^4$} \to
\mbox{$\mathbb{R}\times\mathbf{M}^d$}$, where the d-dimensional
manifold $\mbox{$\mathbf{M}^d$}$ undergoes a topological splitting
into $n$ Einstein spaces: \mbox{$\mathbf{M}^d = \prod_{i=1}^n
\mathbf{M}_i^{d_i}$} , \mbox{$g^{(d)} = \sum_{i=1}^n
g^{(d_i)}_{(i)}$} , \mbox{$R_{\mu \nu}[g^{(d_i)}_{(i)}] =
\lambda_i g^{(d_i)}_{(i)\mu \nu}\, ,\; d = \sum_{i=1}^n d_i$},
with $\lambda_i > 0$. Now, using the conformal transformation
$g^{(d_i)}_{(i)} \to (\lambda_i/\lambda)g^{(d_i)}_{(i)}$, it can
be easily shown \cite{topsplit} that the constituent manifold
$\mbox{$\mathbf{M}^d$}$ is also the Einstein space: $R_{\mu
\nu}[g^{(d)}] = \lambda g^{(d)}_{\mu \nu}$ and the scale factor
$a(\tau)$ satisfies the $(D=d+1)$-dimensional analog of Eq.
(\ref{2.7}):
\begin{eqnarray}\label{6.1} \left(\frac{\dot a}{a}\right)^2 +
\frac{R_d}{d(d-1)} \frac{1}{a^2} -\frac{2}{d(d-1)} \Lambda_D
 -\frac{2}{d(d-1)} \kappa^2_D A\, a^{-\alpha d}\,\,\
= \,\,0,
\end{eqnarray}
where $R[g^{(d)}]=R_d=d\lambda:=d(d-1)k$. Obviously, cosmic
strings correspond to $\alpha = (d-1)/d$, domain walls have
$\alpha = (d-2)/d$, etc. For example, the warp factor of a model
filled with a frustrated network of $(d-2)$-dimensional
topological defects is described in Euclidean region by Eq.
(\ref{4.13}) (in the case of fine tuning $2\kappa^2_DA/d(d-1)
\equiv \bar A^2 =1$ and negative cosmological constant $\Lambda
\equiv 2\Lambda_D /d(d-1)$).

Thus, the brane-world instantons, as well as the birth of the
brane-worlds from them, can be constructed for this
$(d+1)$-dimensional model following a completely parallel
procedure to that was applied to solution (\ref{4.13}). By
instance, in the particular case $\mbox{$\mathbf{M}^d$} =
\textbf{S}^{4} \times \prod_{i=1}^n \textbf{S}^{d_i}$, the
analytic continuation \mbox{$\chi\,\rightarrow \,iHt+{\pi}/{2}$}
with respect to the coordinate $\chi$ of the 4--sphere
$\textbf{S}^{4}$ results in the following Lorentzian metric:
\begin{eqnarray}\label{6.2}ds^2_L = dy^2 + H^2 a^2(y)( - dt^2 +
\frac{1}{H^2}\cosh^2 Ht \, d\Omega^2_{(3)}+ \frac{r_1^2}{H^2}
d\Omega^2_{(d_1)} + \ldots + \frac{r_n^2}{H^2} d\Omega^2_{(d_n)})
,\end{eqnarray}
where $r_i\, , \; i= 1, \ldots ,n\; $ are the radii of the
$d_i$-spheres and $n$ the number of spheres. We thus arrive at a
brane-world model with inflating 4--D part of the brane, plus
frozen, compactified and unobservable (for $r_i \lesssim
10^{-17}cm$) $(d-4)$ dimensions on it. This scenario, with a brane
of codimension one, is of interest because:\\ 1. it points out to
the very interesting possibility for the unification of new
brane-world scenarios with the standard Kaluza--Klein approach.\\
2. it gives a possibility for the localization of the gauge fields
on the brane\footnote{ As it follows from equation (2) of
reference \cite{DRT}, only for $n>0$, where $n$ is the number of
additional compactified dimensions, such localized gauge fields
are normalizable.} \cite{DRT}, \cite{Rubakov}.

\bigskip
{\bf Acknowledgments}

A.Z. acknowledges support by Spanish Ministry of Education,
Culture and Sport (the programme for Sabbatical Stay in Spain) and
the programme SCOPES (Scientific co-operation between Eastern
Europe and Switzerland) of the Swiss National Science Foundation,
project No. 7SUPJ062239. M.B.L. is supported by a grant of the
Spanish Ministry of Science and Technology. This investigation was
supported by GICYT under the research Projects Nos BFM2002-03758
and PB98-0520.

\end{document}